\documentclass{jaa}
\usepackage{natbib}
\usepackage{aas_macros}

\usepackage{graphicx}
\usepackage{hyperref}
\usepackage{caption}

\newcommand{\galex}{{GALEX}}

\def\aap {{A\&A\,}}

\def\apjs {{ApJS\,}}
\def\apjl {{ApJL\,}}

\usepackage{longtable}

\usepackage{lineno}

\usepackage{multirow}

\begin{document}\sloppy
\title{UVIT/AstroSat observation of TW Hya }


\author{Prasanta K. Nayak\textsuperscript{1, 2,*}, Mayank Narang\textsuperscript{2,3}, P. Manoj\textsuperscript{2}, D. K. Ojha\textsuperscript{2}, U. S. Kamath\textsuperscript{4}, Blesson Mathew\textsuperscript{5}, T. Baug\textsuperscript{6}, S. Vig\textsuperscript{7}, S. Chandra\textsuperscript{8}, , G. Maheswar\textsuperscript{4}
}

\affilOne{\textsuperscript{1}Instituto de Astrofísica, Pontificia Universidad Católica de Chile, Av. Vicuña MacKenna 4860, 7820436, Santiago, Chile.\\}
\affilTwo{\textsuperscript{2}Department of Astronomy and Astrophysics, Tata Institute of Fundamental Research, Mumbai, 400005, India.\\}
\affilThree{\textsuperscript{3}Academia Sinica Institute of Astronomy \& Astrophysics, 11F of Astro-Math Bldg., No.1, Sec. 4, Roosevelt Rd., Taipei 10617, Taiwan, R.O.C.\\}
\affilFour{\textsuperscript{4} Indian Institute of Astrophysics, Bangalore, 560034, India.\\}
\affilFive{\textsuperscript{5} Department of Physics and Electronics, CHRIST (Deemed to be University), Bangalore 560029, India. \\}
\affilSix{\textsuperscript{6} S. N. Bose National Centre for Basic Sciences, Sector-III, Salt Lake, Kolkata 700106, India. \\}
\affilSeven{\textsuperscript{7} Indian Institute of Space-science and Technology, Thiruvananthapuram 695547, India. \\ }
\affilEight{\textsuperscript{8} Physical Research Laboratory, Navrangpura, Ahmedabad - 380 009, India. \\}

\twocolumn[{

\maketitle

\corres{nayakphy@gmail.com}


\begin{abstract}
The paper demonstrates the spectroscopic and photometric capabilities of the Ultra-Violet Imaging Telescope (UVIT) to study T-Tauri stars (TTSs). 
We present the first UVIT/Far-UV (FUV) spectrum of a TTS, TW Hya. Based on C~{\sc iv} line luminosity, we estimated accretion luminosity (0.12$\pm$0.03 $L_\odot$) and mass accretion rate (2.4$\pm$0.6 $\times$ $10^{-8} M_\odot /yr$) of TW Hya, and compared these values with the accretion luminosity (0.031$\pm$0.002 $L_\odot$) and mass accretion rate (0.62$\pm$0.04 $\times$ $10^{-8} M_\odot /yr$) derived from spectral energy distribution (SED). 
From the SED, we derive best-fitted parameters for TW Hya: $T_{eff}$ = 3900$\pm$50 K, radius = 1.2$\pm$0.03 $R_\odot$, $\log\, g = 4.0$ and equivalent black-body temperatures corresponding to accretion luminosity as 14100$\pm$25 K. The parameters of TW Hya derived from UVIT observations were found to be matched well with the literature. 
Comparison with International Ultraviolet Explorer (IUE) and Hubble Space Telescope (HST) spectra suggests that UVIT can be used to study the spectroscopic variability of young stars. 
This study proposes leveraging the FUV spectroscopic capabilities of UVIT to contribute to the advancement of upcoming UV spectroscopic missions, including the Indian Spectroscopic Imaging Space Telescope.

\end{abstract}


}]


\doinum{12.3456/s78910-011-012-3}
\artcitid{\#\#\#\#}
\volnum{000}
\year{0000}
\pgrange{1--}
\setcounter{page}{1}
\lp{1}

\section{Introduction}

Study of classical T-Tauri stars (CTTSs) and their accretion processes are important to understand star and
planet formation, as the stars gain a significant fraction of their final mass from circumstellar disks via accretion, and planet formation is also thought to take place during this phase of evolution. 
The CTTSs show signs of enhanced line emission (e.g., Hydrogen Balmer lines, Ca~{\sc ii}) as a signature of active accretion from circumstellar disk \citep{Joy45, Muz98, Stahler_2004fost.book.....S, Hartmann, Alaca}. 
A strong continuum excess long-ward of about 1 $\mu$m in CTTSs indicates the presence of the protoplanetary disk around them. 
Apart from showing strong line emission and IR excess, T-Tauri stars (TTSs) also emit significant excess in the blue and UV continuum \citep{Kuhi, Schneider, nayak_taurus_2024}.

Traditionally, accretion rates onto CTTSs have been derived using the optical emission lines, the blue excess, and veiling in the optical regime \citep[e.g., ][]{ Muzerolle_1998, Rigliaco, Hartmann, Alaca}. However, X-ray and FUV observations offer the most direct method for measuring the accretion process in low-mass stars. 
Material from the disk is channelled onto the stars along the strong magnetic field lines generating strong accretion shock at the surface of the star and producing hot- spots. The post-shock temperature of the infalling material rises to $> 10^6$ K, resulting in the generation of a copious amount of X-ray radiation, which gets partly reprocessed into UV and optical radiation \citep{calvet1998, Schneider_2017}. 
Based on magnetospheric accretion models, most of the accretion funnel has only T $\sim$ 10,000 K \citep{Muzerolle_1998, Kwan_2011}. Therefore, the strong FUV emission lines might be produced in the funnel regions immediately above the shock or from postshock material \citep{ardila2013}. 
FUV spectra of CTTSs are dominated by strong emission lines from neutral and ionized species such as O~{\sc i} $\lambda$1304, Si~{\sc iv} $\lambda \lambda$1394/1403 doublet, C~{\sc iv} $\lambda \lambda$1548/1450 doublet, He~{\sc ii} $\lambda$1640  
\citep{Valenti, Krull, Valenti03, Yang, greg_2023_TWHya}. 
Luminosities of the hot ionized lines are found to correlate with mass accretion rates \citep{Johns-Krull_2000, Yang}. 
\cite{Yang} found a strong correlation between the line luminosity of some of these hot ionized lines and the accretion tracer optical emission lines such as the Balmer lines and the Ca~{\sc ii} K. \cite{Yang} also showed that the FUV continuum luminosity strongly correlates with the accretion luminosity \citep[also see][]{Ingleby11b}. Similar results connecting the UV luminosity to accretion luminosity have also been derived using photometric observation from Ultra-Violet Imaging Telescope \citep[UVIT;][]{nayak_taurus_2024}.

Most of the developments in the UV analysis of CTTSs came with the advent of the International Ultraviolet Explorer (IUE) satellite \citep[IUE:][]{Boggess_1978_IUE_spacecraft, IUE}, which was later followed up by observation from the Hubble Space Telescope (HST). Recently, a large HST UV spectroscopic program was also initiated by the STScI Director in 2019 to study the accretion processes in young high- and low-mass stars in the local universe, named Ultraviolet Legacy Library of Young Stars as Essential Standards \citep[ULLYSES\footnote{https://ullyses.stsci.edu}:][]{ULLYSES_Program_2025}, which was completed in 2023.
Overall, this shows the strength of FUV luminosity as a tool to study the accretion onto pre-main-sequence stars.
In this work, we examine the spectroscopic capabilities of UVIT to study TTSs and discuss whether UVIT can be used to complement the legacy of the ULLYSES program in future. 
\cite{nayak_taurus_2024} recently demonstrated the photometric capabilities of UVIT to study the accretion mechanism in TTSs. However, the spectroscopic capabilities of UVIT to study TTSs have not been explored yet. It has also not been examined whether UVIT can be used to detect the  spectroscopic variability in TTSs. In this paper, we use the UVIT data of TW Hya to explore these aspects.


TW Hya is a CTTS and is the most prominent member of the TW Hya association \citep{Kastner}. Its spectral type ranges from K6/K7 to M2 ($T_\mathrm{eff} = 4200 - 3600$ K) depending on whether measured in blue/optical or near-IR wavelengths \citep{Webb_1999, Yang_2005, Vacca_2011, Debes_2013, McClure_2013,  Venuti}. TW Hya is also the closest T-Tauri star at 60.14 pc \citep{Bailer21}. 
Even though the system is about 5-10 Myr old, there is still IR excess associated with it, indicating the presence of a protoplanetary disk, which has been imaged with HST \citep[e.g.,][]{Debes_2017}, ground-based coronagraphs  \citep[e.g.,][]{Akiyama_2015, Rapson_2015} and ALMA \citep[e.g.,][]{ALMA, Huang_2018} in great detail. The disk structure of TW Hya includes a dust-depleted inner hole, as well as a series of bright rings, with the closest one located at about one astronomical unit (1 AU) from the star. 
These gaps are thought to be created due to the ongoing planet formation in the disk. The relatively old age of the system ($\sim$ 5-10 Myr) for a TTS makes it a candidate for long-lived disks. Understanding the accretion processes in such long-lived disk systems can provide important clues about the fate of planetary systems.  
TW Hya has previously been observed in the UV wavelength with IUE and HST \citep{ Herczeg02, Robinson}. The UV flux from TW Hya is also found to be variable \citep{Robinson, greg_2023_TWHya}, which indicates that the mass accretion from TW Hya is changing with time.

We described observation and data reduction in \autoref{observation}. The results are presented and discussed in \autoref{results}. 
The results are summarized in \autoref{summary}.

\section{Observations and Data reduction}
\label{observation}

\subsection{UVIT/AstroSat}
UVIT, onboard AstroSat satellite, consists of two identical 38 cm coaligned telescopes with a field of view of $\sim$28 arcmin: one is dedicated for the FUV band (120-180nm) and the other for the NUV (200-300nm) and VIS (320-550nm) bands. 
Although UVIT is primarily an imaging instrument, it also has low-resolution slit-less spectroscopic facilities in both NUV and FUV bands: one grating is mounted in the NUV wheel, and two gratings (with orthogonal dispersion axes) in the FUV wheel. The advantage of the UVIT is that the telescope simultaneously observes the targets in both the FUV and NUV bands. This helps to better constrain the UV spectra of TTSs. However, the NUV channel is currently inoperative and only the FUV channel remains operative.
The maximum efficiency is achieved in the $-$2 order of the FUV gratings with a spectral resolution of 15 \AA~\citep{tandon2020, Dewangan}.

\begin{figure*}[h!]
\centering
\includegraphics[width=0.8\linewidth]{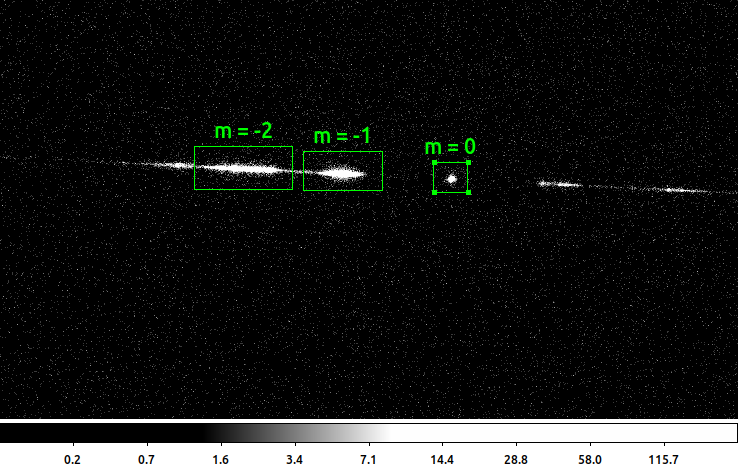}
\includegraphics[width=0.3\linewidth]{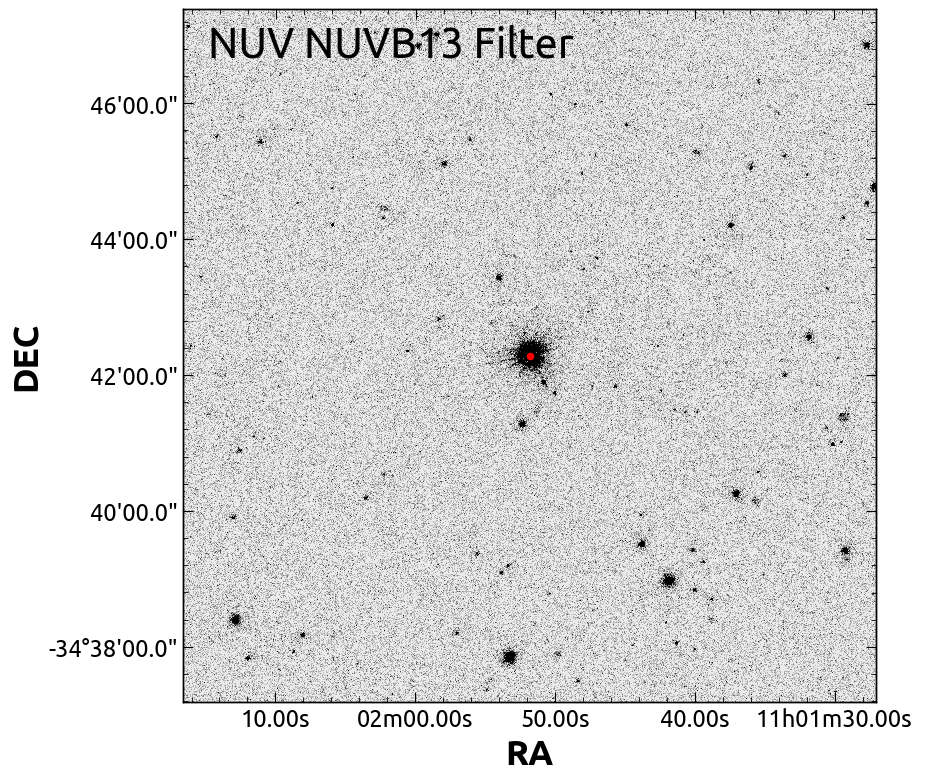}
\includegraphics[width=0.3\linewidth]{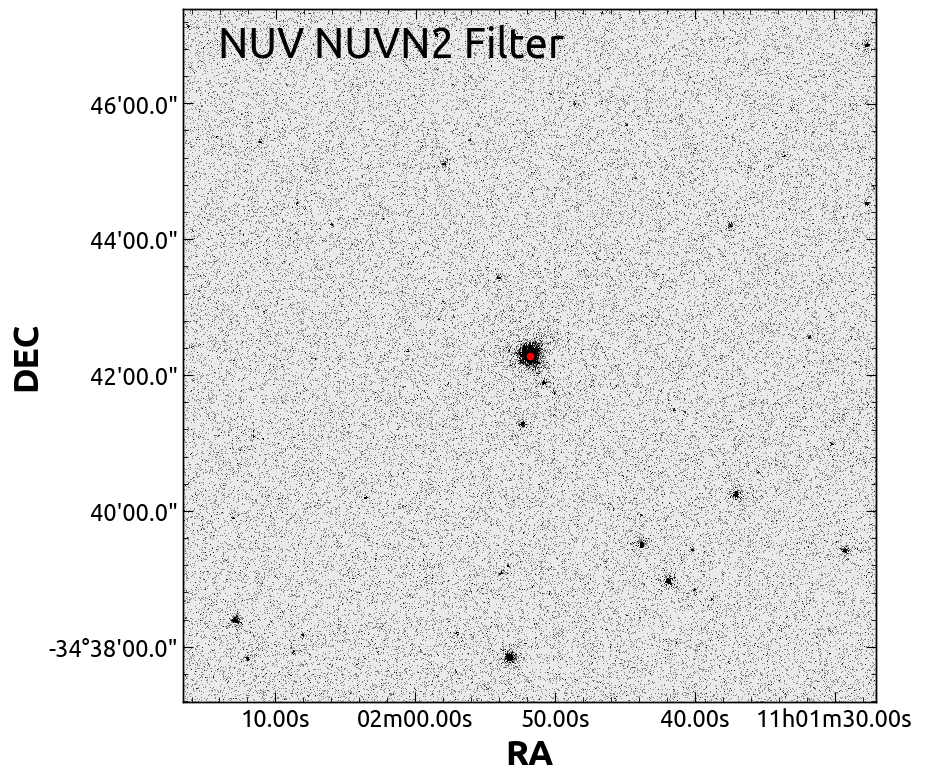}
\includegraphics[width=0.3\linewidth]{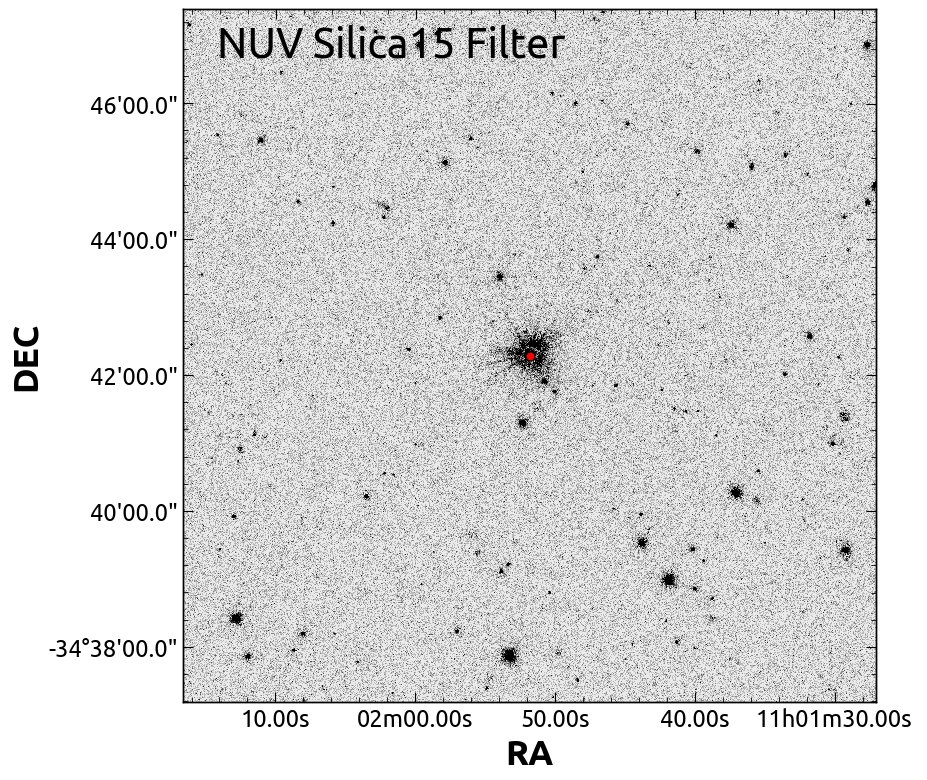}
\caption{(top) FUV Grating-1 image of TW Hya. (bottom ) NUV images of TW Hya in different UVIT filters. 
} 
\label{spec_image}
\end{figure*}

The UVIT observations of TW Hya were carried out as part of the GT proposal (ID: G07\_086, PI: D. K. Ojha) on 2017 April 07. The FUV observation was performed in the spectroscopic mode, while the NUV observations were completed in the photometric mode in a wide (N242W/Silica-1), a medium (N245M/NUVB13), and a narrow (N279N/NUVN2) band. The level-1 raw data were corrected for spacecraft drift, flat-field and distortion, and the orbit-wise images were generated using the software CCDLAB \citep{postma2017}. Then, the orbit-wise images were co-aligned and combined to generate final science-ready images in each filter. Astrometry was also performed using CCDLAB by comparing the $Gaia$-DR3 source catalogue. The details about the telescope and the instruments are available in \citet{subra2016, tandon2017b}, and the instrument calibration can be found in \cite{tandon2017a,tandon2020}.

The FUV and NUV observations were taken simultaneously and completed over several orbits to obtain a total exposure times of $\sim$10 ksec in each telescope. The NUV photometric observations are completed in the three bands as follows: N245W (5197 s), N279N (2951 s), and N242W (1845 s). However, in the case of FUV Grating 1 observation, we were able to retrieve the data for an exposure time of 8062 s. We lost 2385 s of data due to large distortion, where {\sc CCDLAB} is unable to correct them properly. The observational details with filter name, wavelength range, date and start time of the observations, and the corresponding exposure times in those filters are shown in \autoref{UVIT_obs}. In the case of spectroscopic observations, we have mentioned the exposure times and date of observations for each orbit. We provided this additional information to demonstrate that though the total observation time for FUV observation is $\sim$8 ksec (2.2 hours), the observation was carried over a period of $>$ 10 hours due to a gap in observation between two orbits. This information is used in \autoref{spec_variability} to check if UVIT can be used to detect hourly variation in accretion rate.

\begin{table*}[h!]
\centering
\caption{The AstroSat/UVIT observations log with the filter range and exposure time. }
\label{UVIT_obs}
\footnotesize
\begin{tabular}{cccccc}
\hline
 Filter/ & Wavelength range  &   observation date    & start time & exposure time & SNR \\ 
 Spectrograph &  (\AA)    & (UTC)  & (UTC) & (s) & \\
\hline 
\hline
  &   \multicolumn{3}{c}{ Details of UVIT observation}    &  & \\ \hline
NUV Silica-1 (N242W) & 1700.00 - 3050.00 &  2017-04-06  & 21:24:23.04  & 1844.698 & \\
  &     &  & \\
NUVN2 (N279N) & 2722.26 - 2877.19 &  2017-04-07  &  01:44:15.55  & 3036.285 & \\
  &   &   \\
NUVB13 (N245M) & 2195.08 - 2634.78 &   2017-04-07  &  08:26:02.59  & 5202.075  & \\
  &   &    \\
FUV Grating 1 & 1200.00 - 1800.00 &  2017-04-07  & 02:16:38.91 & 469.343   & 3.5\\
&   &  2017-04-07  & 03:30:13.64 & 768.007   & 4.3 \\
&   &  2017-04-07  & 05:12:43.67 & 1039.193  & 5.0\\
&   &  2017-04-07  & 06:54:43.41 & 1337.846  & 5.8\\
&   &  2017-04-07  & 08:37:13.62 & 1634.716  & 6.5\\
&   &  2017-04-07  & 10:19:42.82 & 1879.611  & 6.6\\
&   &  2017-04-07  & 12:01:17.21 & 932.879   & 4.3\\
\cline{2-6}
  &     \multicolumn{3}{c}{Total exposure time and SNR of Grating 1 =}   & 8061.595 & 14.0 \\ \hline \hline 
  &   \multicolumn{3}{c}{Observational details of IUE spectra}    &  & \\ \hline
  Short Wavelength Prime & 1150.00 - 2000.00 & 1979-10-29$^1$ & 11:18:17 & 9119.614 & 17.9 \\
  Short Wavelength Prime & 1150.00 - 2000.00 & 1984-07-16$^2$ & 18:16:33 & 5999.021 & 6.5 \\ \hline \hline
    &   \multicolumn{3}{c}{Observational details of HST spectra}    &  & \\ \hline 
  COS/FUV: G160M$^3$ & 1360.00 - 1775.00 & 2021-04-03 &  04:28:45.4 & 30 & 2.0 \\ 
     &  & 2021-04-05 &  04:23:51.3 & 30 & 1.9 \\
     &  & 2021-04-08 &  13:09:56.4 & 30 & 1.9 \\
     \hline
\end{tabular}%
\caption*{ {\footnotesize \textbf{Note.}  $^1$Data ID: SWP07036. $^2$Data ID: SWP23471. $^3$Proposal ID: 16109}}
\end{table*}

For the photometric data reduction, we used the DAOPHOT tasks and packages in the Image Reduction and Analysis Facility (IRAF) software \citep{stetson1987}. 
As the observed field is not crowded, we performed aperture photometry and applied aperture corrections to aperture magnitudes. The final magnitudes are calculated in the AB magnitude system by adding zero-point magnitudes for the corresponding bands, obtained from \citet{tandon2020}. In the case of slit-less spectroscopic observation, we used m=$-$2 order, as shown in Figure \ref{spec_image}, for further analysis. The wavelength and flux calibrated spectrum is generated using the pipeline by \citet{Dewangan}. We achieve an SNR of 14 for this spectrum. The spectrum is shown in the top panel of  \autoref{UVSpec}.

\subsection{IUE}
\begin{figure*}[h!]
\centering
\includegraphics[width=0.7\linewidth]{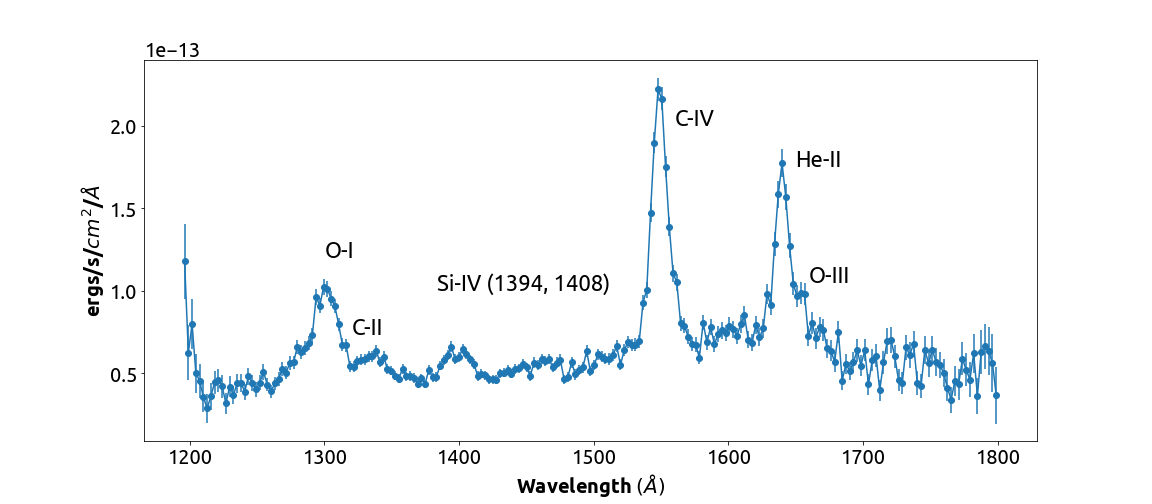}
\includegraphics[width=0.6\linewidth]{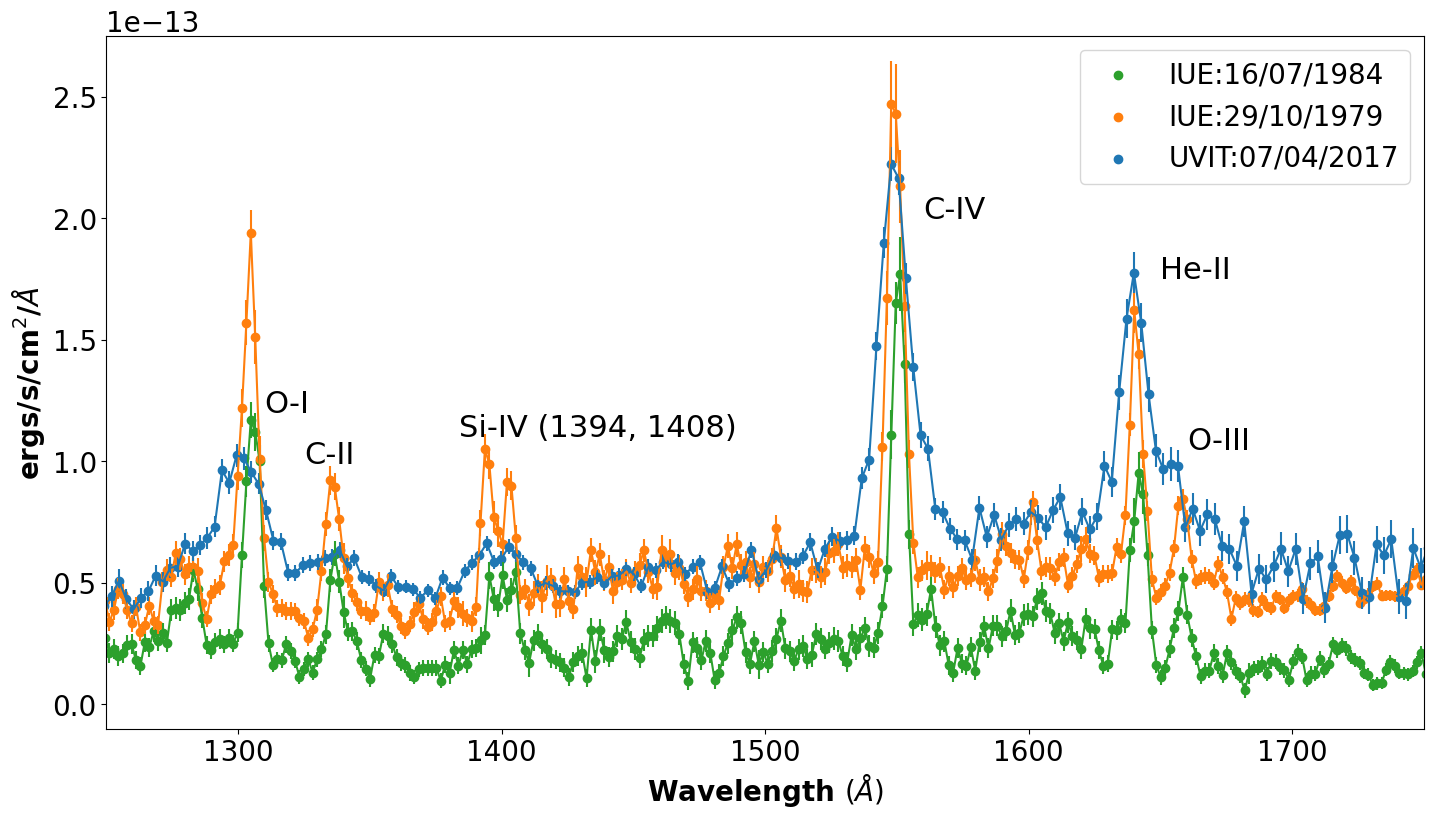}
\includegraphics[width=0.6\linewidth]{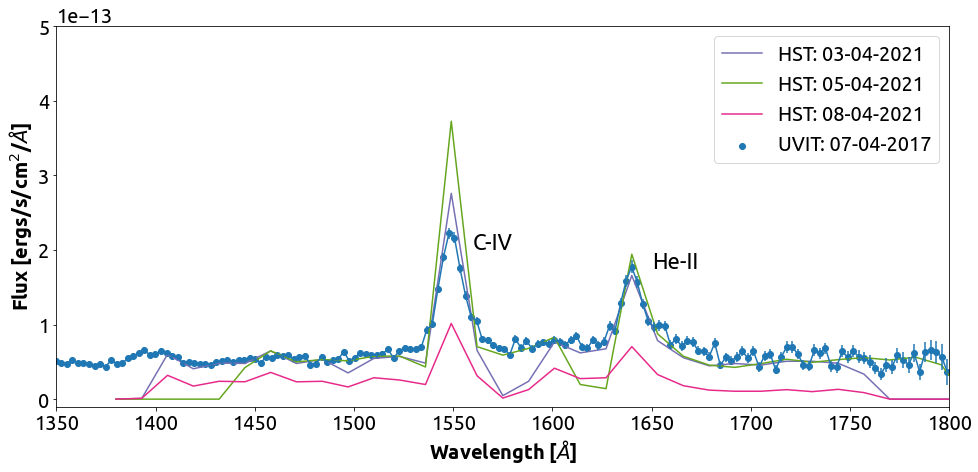}
\caption{(Top panel): UVIT/FUV spectrum of TW Hya. (Middle panel): UVIT spectrum is compared with the low-resolution IUE spectra of two epochs. (Bottom panel): UVIT spectrum is compared with the HST spectra (from the ULLYSES archive) of three epochs, where HST spectra are degraded to UVIT resolution. } 
\label{UVSpec}
\end{figure*}

We used low resolution spectroscopic observations obtained from the IUE satellite to compare the UVIT observation. We found two epochs of  reduced FUV spectra\footnote{https://archive.stsci.edu/iue/}, observed with the short wavelength prime (SWP) camera operated in a low-dispersion spectral mode with spectral resolution of $\sim 6\AA$. The IUE data were reduced with the latest IUE New Spectroscopic Image Processing System  \citep[NEWSIPS:][]{Nichols_and_Linsky_1996_NEWSIPS} software, which produces wavelength and flux calibrated spectra with $\sim$10\% uncertainty in flux calibration, 
while the uncertainly between multi-epochs of spectra can be up to $\sim$10--15\% due to time-dependent systematic effects \citep{Massa_and_Fitzpatrick_2000_NEWSIPS}.
The observational details along with SNR of the spectra are tabulated in \autoref{UVIT_obs}. The data ID for the two observations used in this study are SWP07036 and SWP23471, with exposure times $\sim$9120 s and $\sim$6000 s, respectively. The IUE spectra are compared with the UVIT spectrum in the middle panel of \autoref{UVSpec}.


\subsection{HST/COS}

We also used high-resolution ($R \sim$11,000) FUV spectra obtained using the G160M grating on the Cosmic Origins Spectrograph (COS) aboard the HST \citep{Kerman_2023_HST-COS_resolution}. The observations were carried out as part of the ULLYSES program. We utilized three epochs of pipeline reduced data (HST Proposal ID: 16109) downloaded from the ULLYSES archive\footnote{https://ullyses.stsci.edu/ullyses-download.html}. The data were processed using the {\it CalCOS} reduction pipeline. The absolute flux calibration uncertainty of COS/FUV spectra is typically $\leq$5\%, while the relative flux stability over the multiple epochs after time-dependent sensitivity corrections remains within $\leq$2\% \citep{Fischer_2018_cos_flux_cal, Miller_2024_cos_flux_cal}. The HST spectra degraded to UVIT resolution are compared with UVIT spectrum in the bottom panel of \autoref{UVSpec}.

\section{Results and Discussion}
\label{results}

\subsection{FUV spectrum of TW Hya}

In the top panel of \autoref{UVSpec}, we have shown the FUV/UVIT spectrum of TW Hya. 
The spectrum clearly shows the detection of several strong lines such as  O~{\sc i} $\lambda$1304, the Si~{\sc iv} $\lambda \lambda$1394/1408 doublet, the C~{\sc iv} $\lambda$1549 doublet, and the He~{\sc ii} $\lambda$1640. We have also compared the spectrum with low-resolution IUE spectra and high-resolution HST spectra obtained from the archive in the middle and bottom panels of Figure \ref{UVSpec}, respectively. Before comparing with the HST spectra, we degraded its spectral resolution to UVIT resolution using {\it coronagraph} package \citep{Robinson_2016_coronagraph, Lustig_2019_coronagraph}. The epochs of these observations are also mentioned in the figure. The FUV/UVIT spectrum matches well with both IUE and HST spectra in terms of line detection; however, due to relatively lower resolution \citep[$\sim$15 \AA; ][]{Dewangan} than IUE, the FUV/UVIT spectrum is unable to resolve the  O~{\sc iii} emission.  
The multi-epoch spectra from both IUE and HST show the signature of variation in the continuum as well as line flux. The UVIT spectrum is found to match nicely with one of the epochs for both IUE and HST. Part of this variability could be attributed to the variable accretion rate. However, some of the variability could be due to uncertainty in the flux calibration.  
Multiple IUE observations of a given target revealed that the continuum flux varies by up to 10\% across different epochs \citep{Bohlin2018}. 
The variability observed ($\sim$ factor of two) between the two IUE epochs of TW Hya surpasses the known flux offset. 
While UVIT exhibits higher flux uncertainty \citep[$\sim$21-25\%;][]{Dewangan}{}{}, it remains less than the flux variation observed between the two IUE epochs.
This shows a clear signature of accretion variability of TW Hya and indicates that UVIT FUV spectra can be used to detect the variable accretion despite of relatively large uncertainty in continuum flux. In Table \ref{table_line_flux}, we list the line fluxes (O~{\sc i}, C~{\sc iv}, and He~{\sc ii}) in TW Hya spectra for UVIT, IUE and HST observations. We determined line fluxes by deducting the continuum, estimated using a linear fit with wavelength, and then integrating the residual emission within the line. The uncertainties in the estimated line fluxes were computed by combining the observational errors in the spectra with the systematic uncertainties associated with the absolute flux calibration. We adopted calibration uncertainties of 25\% for UVIT, 10\% for IUE, and 5\% for HST.

\begin{table*}[h!]
\footnotesize
\centering
\caption{The line flux measured from the UVIT and archival IUE spectra.}
\begin{tabular}{|cc|ccc|ccc|}
\hline
\multicolumn{2}{|c|}{} & \multicolumn{6}{c|}{Line Flux ($\times 10^{-13}\, ergs/s/cm^{2}$)} \\ \cline{3-8} 
\multicolumn{2}{|c|}{} & \multicolumn{1}{c|}{UVIT} & \multicolumn{2}{c|}{IUE} & \multicolumn{3}{c|}{HST}  \\ \hline
\multicolumn{1}{|c|}{Line} & Wavelength & \multicolumn{1}{c|}{07/04/2017} & \multicolumn{1}{c|}{29/10/1979} & 16/07/1984 & \multicolumn{1}{c|}{03/04/2021} & \multicolumn{1}{c|}{05/04/2021} & 08/04/2021 \\ \hline
\multicolumn{1}{|c|}{O~{\sc i}} & 1302.168 & \multicolumn{1}{c|}{8.0 $\pm$ 2.0} & \multicolumn{1}{c|}{10.9 $\pm$ 1.2} & 7.3 $\pm$ 0.8 & \multicolumn{1}{c|}{--} & \multicolumn{1}{c|}{--} & -- \\ \hline
\multicolumn{1}{|c|}{C~{\sc iv} doublet} & 1548.187 - 1550.772 & \multicolumn{1}{c|}{23.5 $\pm$ 6.0} & \multicolumn{1}{c|}{15.1 $\pm$ 1.6} & 9.9 $\pm$ 1.1  & \multicolumn{1}{c|}{30.1 $\pm$ 1.6} & \multicolumn{1}{c|}{43.6 $\pm$ 2.2} &  10.6 $\pm$ 0.6 \\ \hline
\multicolumn{1}{|c|}{He II} & 1640.420 & \multicolumn{1}{c|}{11.7 $\pm$ 3.0} & \multicolumn{1}{c|}{6.2 $\pm$ 0.7} & 4.4 $\pm$ 0.5   & \multicolumn{1}{c|}{14.1 $\pm$ 0.8} & \multicolumn{1}{c|}{16.8 $\pm$ 1.0} &  {6.1 $\pm$ 0.4} \\ \hline
\end{tabular}%
\caption*{{\footnotesize Note. The absolute flux calibration uncertainties (25\% for UVIT, 10\% for IUE, and 5\% for HST) are included in the errors estimations of line-fluxes.} 
}
\label{table_line_flux}
\end{table*}

\subsection{Mass accretion rates from FUV spectroscopy}

The C~{\sc iv} line present in the FUV spectrum is an accretion tracer and can be used to compute the mass accretion rate onto TTSs. The C~{\sc iv} line also correlates with the accretion luminosity as well as FUV luminosity. Following \cite{Yang} we can express the line luminosity of the C~{\sc iv} $\lambda$ 1549 line as a function of accretion luminosity:  

\begin{equation}
    log(L_\mathrm{C~{IV}}) = -2.766 + 0.877 \times log(L_\mathrm{acc}) 
    \label{eq1}
\end{equation}

Using the line fluxes listed in Table \ref{table_line_flux}, we compute the $L_\mathrm{acc}$ from the UVIT and IUE spectra. We derive an $L_\mathrm{acc}$ of 0.12$\pm$0.03 $L_\odot$ from the UVIT spectrum while from the IUE spectra, we derive an $L_\mathrm{acc}$ between 0.04 -- 0.07 ($\pm$0.01) $L_\odot$ and from the HST spectra $L_\mathrm{acc}$ lies between 0.05 -- 0.24 ($\pm$0.01) $L_\odot$. We notice that the value of $L_\mathrm{acc}$ from UVIT observation matches closely with the $L_\mathrm{acc}$ range observed in IUE. HST observation shows relatively larger range in $L_\mathrm{acc}$, which also matches with UVIT and IUE observations.

The accretion luminosity $L_\mathrm{acc}$ can then be converted into the mass accretion rate following \cite{gullbring1998} : 

\begin{equation}
    L_\mathrm{acc}  =  \frac{G\,M_* \dot{M}_\mathrm{acc}}{R_*}  \left( 1 - \frac{R_*}{R_{in}} \right)
    \label{Lacc_Macc_rel}
\end{equation}

where $M_*$ and $R_*$ are the mass and radius of the star, $\dot{M}_\mathrm{acc}$ is the mass accretion rate and $R_{in}$ is the dust truncation radius. The typical value for $R_{in}$ is $R_{in}=  5 R_*$ and for  $R_*/M_*\sim 5 R_\odot/M_\odot$ \citep{Muz98}. Using  Equation \ref{Lacc_Macc_rel}, we derive the mass accretion rate onto TW Hya to be between 0.8--4.8 $\times 10^{-8}$~$M_\odot/yr$ (2.4$\pm$0.6 $\times 10^{-8}$~$M_\odot/yr$ from UVIT observation; 0.8--1.4$\pm$0.2 $\times 10^{-8}$~$M_\odot/yr$ from IUE and 1.0--4.8$\pm$0.2 $\times 10^{-8}$~$M_\odot/yr$ from HST). Similar values for mass accretion rates have been derived for TW Hya using various other techniques \citep[e.g.,][]{Kastner, Venuti, Robinson, greg_2023_TWHya, Wendeborn_2024_TWHya_ULLYSES}.

\subsection{Mass accretion rates from SED analysis}
\label{sed_analysis}

The accretion process in TTSs causes excess radiation in the UV region and the accretion luminosity peaks in the FUV region with equivalent black body temperature $\sim$10$^4$ K \citep{calvet1998}. Simultaneous FUV and NUV observation provide a better estimation of UV luminosity (due to accretion or chromospheric activity) by constructing spectral energy distribution (SED) and measuring excess emission over photospheric luminosity. Simultaneous FUV and NUV observations also help us to estimate the equivalent temperature corresponding to UV luminosity by comparing the excess UV luminosity with a black body SED \citep{nayak_taurus_2024}. 
As TW Hya is observed simultaneously in FUV and NUV filters, we tried to estimate the accretion luminosity and its corresponding peak temperature by performing a two-component model spectra that fit the UV and optical part of the SED of TW Hya.


\begin{figure*}
\centering
\includegraphics[width=0.48\linewidth]{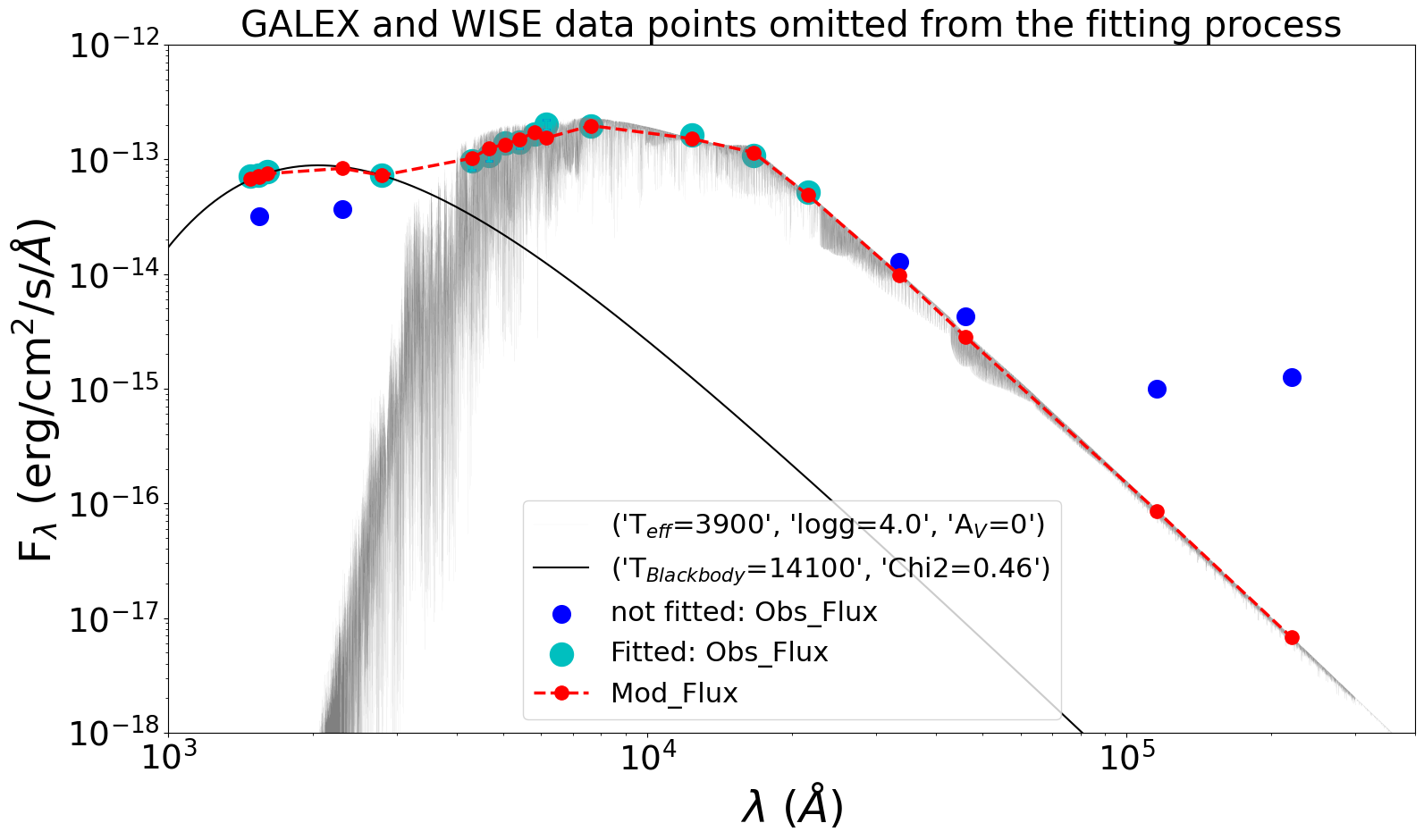}
\includegraphics[width=0.48\linewidth]{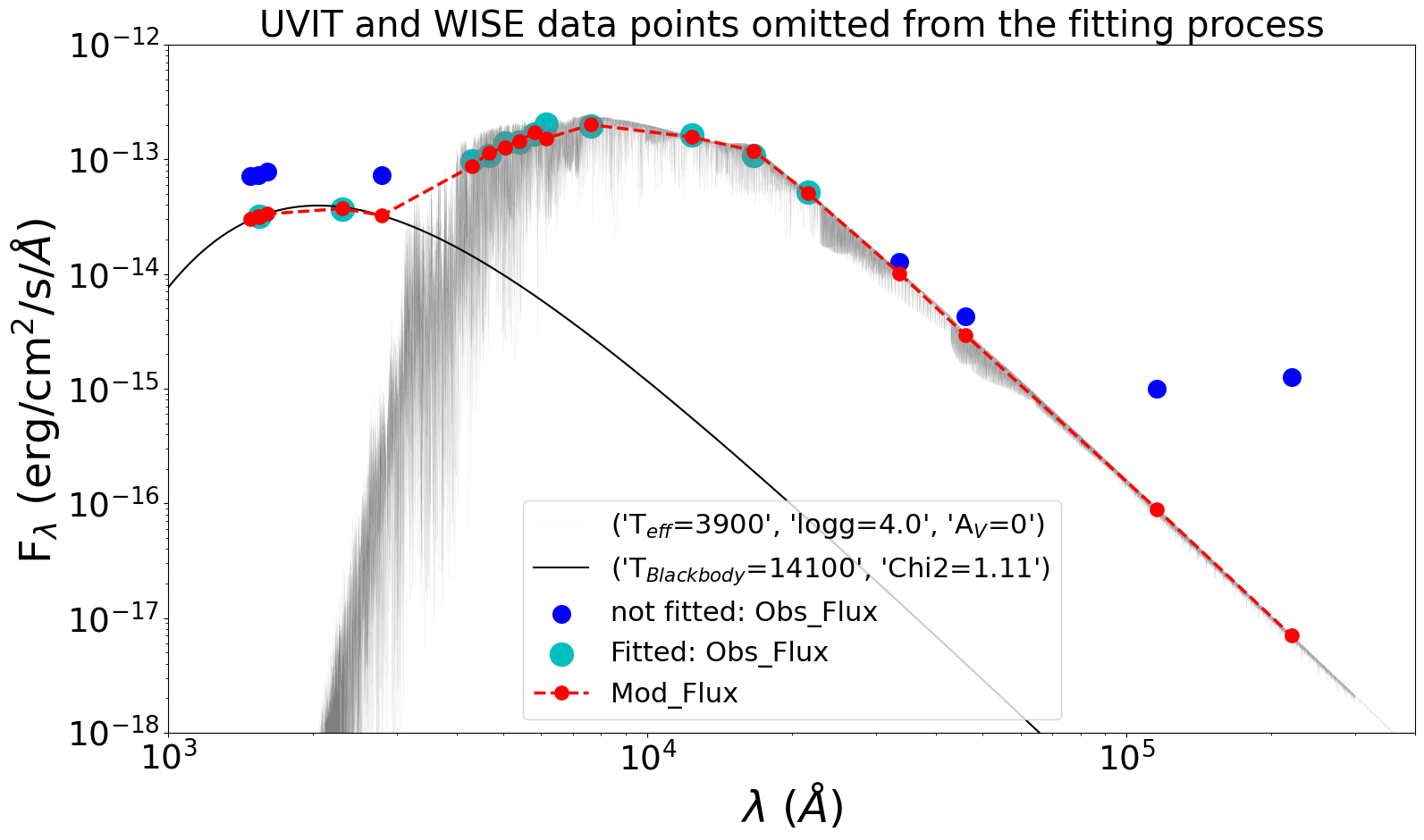}
\caption{SEDs of TW Hya are shown here. The grey and black lines represent the best-fitted synthetic black body and dwarf spectra, respectively, on observed fluxes (cyan and blue points). Cyan points are included in the fitting algorithm, while the blue points are avoided. The red dashed line indicates the expected combined model flux from the fitted synthetic spectra. (left), We omitted GALEX and WISE data in the fitting process. (right), We omitted UVIT and WISE data. 
The best-fitted parameters are mentioned in the legends.  } 
\label{UVSED}
\end{figure*}


The TW Hya is detected in all three NUV/UVIT filters; however, it gets saturated in the NUV~Silica-1 and NUVB13 filters. Therefore, we used NUV photometry only in the NUVN2 band. 
To get the FUV magnitudes in different UVIT filters, we convolved the FUV spectrum with the corresponding filter response curve and obtained FUV magnitudes in F148W, F154W, and F169M bands. 
To construct the SED, we have included available photometric data from the archive along with AstroSat/UVIT data points, which are \galex\ GR6+7 \citep[][]{bianchi2017}, $Gaia$-DR3 \citep{Gaia_eDR3_2021, Gaia_DR3_catalog_validation}, APASS DR9 \citep{apass}, 2MASS \citep{2mass}, and WISE \citep{allwise}. 
We have used virtual observatory (VO) tool, VOSA (VO SED Analyzer; \citet{bayo2008}) developed by Spanish Virtual Observatory to generate SEDs 
and to fit theoretical model spectra of a black body and a dwarf star \citep[BT-Settl-CIFIST model;][]{BTSettl2011} in the UV and optical regions of the observed flux distribution. We did not include WISE data points in IR regions for the SED fit as a substantial excess emission in the near-IR region comes from the disk.  
We have used $\log\, g$ values between 4 to 5 and full T$_{eff}$ range of 1200 to 7000 K with a resolution of 100 K for the BT-Settl-CIFIST model. In the case of black body, 
we used spectra for a range of 5000 to 20000 K with a resolution of 50 K. We also kept $A_\mathrm{v}$ as a free parameter in the fitting process with a given range of 0 to 5.

From the literature, we know that TW Hya has a variable accretion rate \citep{Robinson, greg_2023_TWHya}. As the signature of accretion variability is more prominent in UV compared to optical observations, and also UV observations from GALEX and UVIT spanning two different epochs, we attempted to fit the observed SED twice, separately including UVIT and GALEX data. First, we include only UVIT data points in the UV region for the fitting (left panel of \autoref{UVSED}), and second, we include only the GALEX data points in the UV region \autoref{UVSED}. 
The observed flux measurements are denoted as cyan and blue points, where cyan points fall in the UV and optical region of the energy distribution which are included in the dual component fitting of model spectra. The blue points are not included in the fitting process. The black and grey lines represent the best-fitted black-body and BT-Settl-CIFIST spectra, respectively to the cyan points. The combined fluxes from these two model spectra in different wavebands are denoted in red and connected with the red dotted line. The overlap between red and cyan points indicates the goodness of the dual fitting. The values of reduced $\chi^2$, temperatures, $\log\, g$ corresponding to the best-fitted spectra are also noted in the legends.

We notice that VOSA provides very similar stellar parameters for both the fittings but with a relatively large deviation in UV or black-body luminosity. 
From both the SEDs, we derive the best fit $T_\mathrm{eff}$ of 3900$\pm$50 K, radius of 1.2$\pm$0.03 $R_\odot$ and $\log\, g = 4.0$ for TW Hya. This is similar to the $T_\mathrm{eff}$ reported in the literature for the star \citep{Venuti}. 
We notice a small deviation in optical luminosity (i.e. the luminosity of TW Hya) between two different fittings: 0.30 L$_\odot$ for the SED with UVIT data and 0.31 L$_\odot$ for the SED with GALEX. 
While TW Hya shows similar optical luminosity over two epochs, the UV luminosity has a significant variation. The UV luminosity (i.e. accretion luminosity,  $L_\mathrm{acc}$) is more than twice in the case of UVIT data (0.031$\pm$0.002 L$_\odot$) compared to that for GALEX data (0.014$\pm$0.002 L$_\odot$). 
However, black-body temperatures from both the SEDs are found to be the same as 14100$\pm$25 K. This range in $L_\mathrm{acc}$ value matches well 
with the literature \citep[see Table 7 in][]{greg_2023_TWHya}.   
This finding not only confirms the presence of accretion variability in TW Hya but also demonstrates the importance of SED analyses to probe the accretion variability in TTSs. 
The best fitted SED also provides the value of $A_\mathrm{v}$ = 0 which supports the results of \cite{Herczeg04} and also with the Gaia catalog \citep{Gaia_eDR3_2021}.

Comparing $L_\mathrm{acc}$ value with that obtained from C~{\sc iv} line flux of UVIT spectra, we notice that latter value is $\sim$3 times higher compared to the former. Variability in accretion rate might not be the reason for getting higher $L_\mathrm{acc}$ value from C~{\sc iv} line flux, as NUV photometry and FUV Grating 1 observations are performed simultaneously. 
The difference in $L_\mathrm{acc}$ which also reflects in accretion rate estimation, could be due to a large systematic error in the flux calibration in the UVIT FUV spectrum. As mentioned in the calibration paper \citep{Dewangan}, the flux density in FUV broadband filters is always found to be 21-25\% lower compared to that measured with the FUV gratings. The discrepancy is at a level of 2.5$\sigma$. \citet{Sukrit_2023_HIP23309} studied the UVIT FUV spectrum of a M-dwarf HIP 23309 and found that the flux is overestimated by 40\% compared to HST STIS spectra. So, it could be an effect of a large systematic error in flux calibration. Due to the lower resolution and a large systematic error in the UVIT FUV Grating calibration, we are overestimating the value of $L_\mathrm{acc}$. In contrast, the $L_\mathrm{acc}$ obtained from the SED fit is constrained by both FUV and NUV, providing more direct measurements of it. 
Another contribution might come from the errors associated with \autoref{eq1}, the relation between C~{\sc iv} line luminosity and $L_\mathrm{acc}$, given by \citet{Yang}. The relation is obtained from the distribution with a median scatter of 0.60 dex.

We further estimated the mass accretion rate from accretion luminosity using the \autoref{Lacc_Macc_rel}. We find the accretion rate of TW Hya as 6.2 $\times 10^{-9}$~$M_\odot/yr$ (for the SED fit with UVIT data) and 2.8 $\times 10^{-9}$~$M_\odot/yr$ (for the SED fit with GALEX data). The values match well with our spectroscopic analysis, as well as 
with the literature values \citep{Venuti, Robinson, greg_2023_TWHya, Wendeborn_2024_TWHya_ULLYSES}. 
The VOSA tool also provides us the values of radii corresponding to the best-fitted black-body and photospheric (BT-Settl-CIFIST) models. 
We have used this information to compute the filling factor (area of the UV emitting region or area covered by the accretion hot spot/ area of the star) to be 0.18\% (for the SED fit with UVIT data) and 0.08\% (for the SED fit with GALEX data). This is comparable to the filling factor derived for TW Hya of 0.26\% by \cite{ingleby2013}. 
However, variations in the filling factor lead to increase or decrease in the shock spectrum regardless of wavelength, essentially scaling the luminosity of the emission \citep{ingleby2013}. As we have seen here, filling factor from the UVIT observation is found to be $\sim$2 times larger compared to that from the GALEX observations, and the same scenario reflects in the accretion (black-body) luminosity as well.

\begin{figure*}[h!]
\centering
\includegraphics[width=0.45\linewidth]{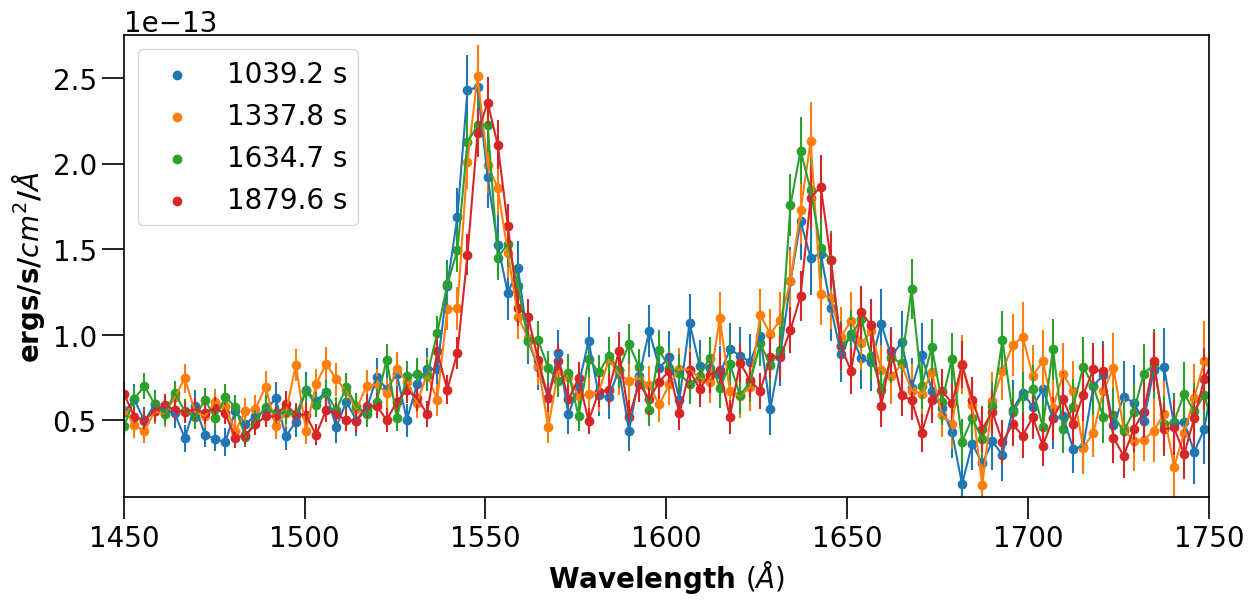}
\includegraphics[width=0.45\linewidth]{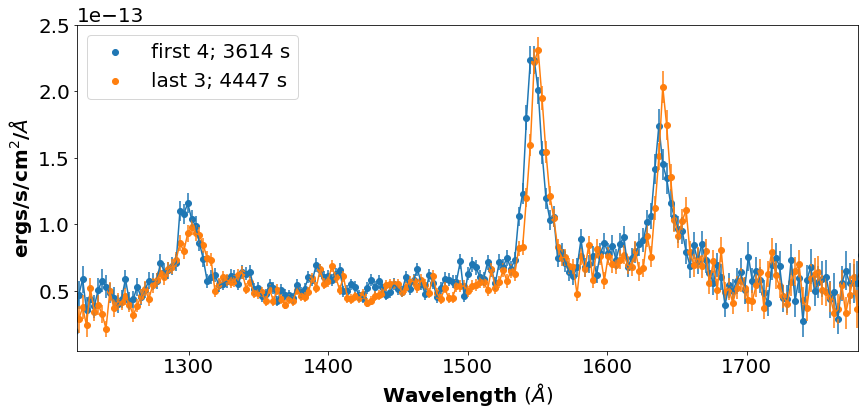}
\includegraphics[width=0.45\linewidth]{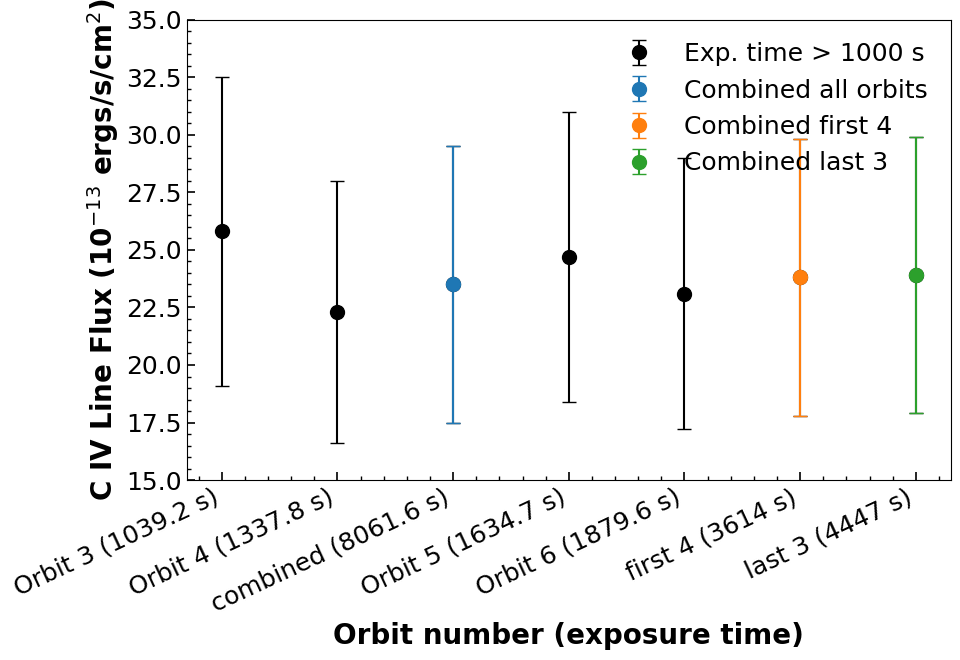}
\includegraphics[width=0.45\linewidth]{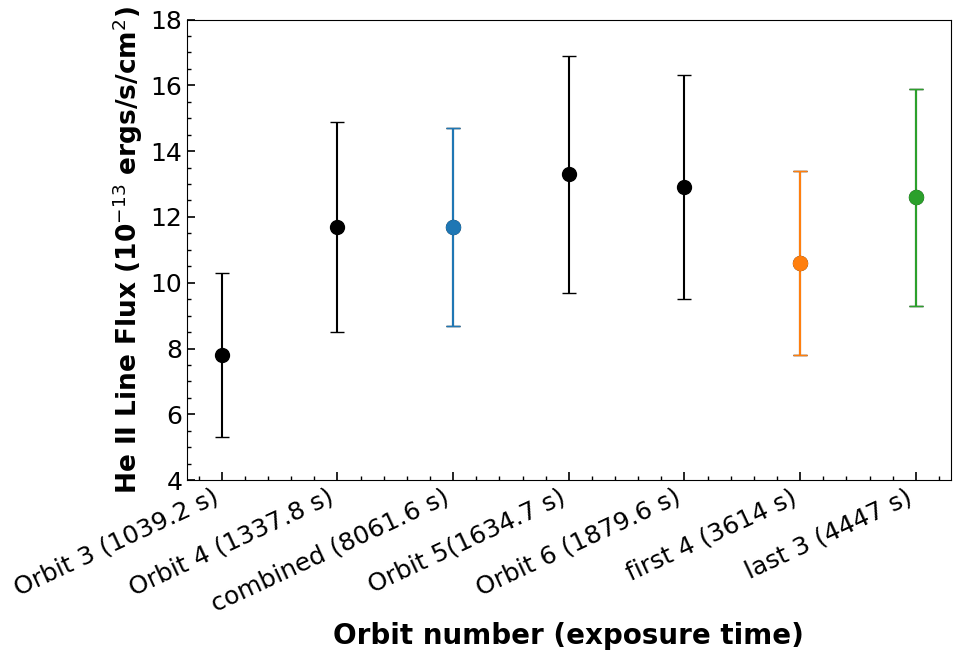}
\caption{UVIT/FUV spectrum and line fluxes (C~{\sc iv} \& He~{\sc ii}) of TW Hya for different orbits. {\it(Top left)} The spectra of different orbits with exposure times $>$1000s are compared. {\it(Top right)} Comparison between the combined spectra from the first four orbits and the last three orbits. Exposure times are mentioned in the legend. {\it(Bottom left)} C~{\sc iv} line flux comparison between the above mentioned four individual orbits with exposures times $>$1000 s (black points), fully combined spectrum of all the orbits (blue point) and the partially combined spectra for first four (orange) and last three (green) orbits. The blue shaded region represents the 1$\sigma$ error in the line flux for the fully combined spectrum. The comparison plot suggests that the observed variations lie within the 1$\sigma$ uncertainly of the fully combined spectrum. {\it(Bottom right)} He~{\sc ii} line flux comparison for different orbits similar to bottom left panel. } 
\label{UVSpec_different_orbits}
\end{figure*}

\subsection{The prospect to study accretion variability using UVIT FUV spectra} \label{spec_variability}
In this section, we try to explore the capability of the UVIT to detect the intra-day variability in the line luminosity. As the UVIT observations are done in photon-counting mode, we have information about the photons in every sub-second (a maximum rate of $\sim$29 frames/s), and we have the leverage to subdivide the total exposure time at the required time-scale bin. In this way, we can look for any observed variability in line luminosity on an hourly basis. From \autoref{UVIT_obs}, we notice that between every orbit observation, there is a gap of $\sim$1 hour. Due to low-resolution, we decided to include the observations orbit-wise and consider those orbits which have exposure times greater than 1000 seconds to achieve an SNR of 5 or more. This also provides us a time cadence of $\sim$1 hour. In the top panel of \autoref{UVSpec_different_orbits}, we compared the spectrum from four orbits, focusing on C~{\sc iv} and He~{\sc ii} lines. Visual inspection suggests that there are no variation in the line luminosities among the spectra, and the errors are significantly high. The observed shift in the spectra of different orbits along the wavelength is  mostly due to imperfect alignment during the combination of multiple observational frames and is within the UVIT's spectral resolution of 15 \AA~\citep{tandon2020, Dewangan}. Individual spectrum for all the seven orbits are shown in  \autoref{diff_orbit_spectrum} in \ref{appendix_orbit_wise_spectrum}. The start time of observation and exposure time for each orbit is mentioned in the legend. 

Next, we divided the total observation into two segments by grouping the first four and the last three orbits (see \autoref{UVIT_obs}), and compared them in the upper right panel of \autoref{UVSpec_different_orbits}. We obtained SNRs of $\sim$10 and $\sim$11 for these two observations. Here, we also notice that the observed variation in the line luminosity or the continuum flux remain within the measurement uncertainties. 
To further quantify possible variability in line flux and assess whether accretion variability can be detected on a cadence of $\sim$1 hour, we measured the C~{\sc iv} and He~{\sc ii} (prominent lines in the spectrum) line flux for the four individual orbits with exposure times $>$1000 s. These were then compared with the line flux derived from the fully combined spectrum (total exposure time $\sim$ 8 ks), shown in the bottom panels of \autoref{UVSpec_different_orbits}. 
The blue point represent the line flux from the fully combined spectrum, with shaded region indicating its 1$\sigma$ uncertainty. Black points correspond to the individual orbit measurements with exposure times labeled on the x-axis. The orange and green points denote the partially combined spectra from the first four and last three orbits, respectively. All measurements lie within the 1$\sigma$ uncertainty of the fully combined spectrum. 
The uncertainty also includes the systematic error associated with the absolute flux calibration (25\%). The comparison indicates that TW Hya did not undergo any significant change in accretion rate during the observation that could be detected with the low-resolution UVIT spectra.

We further notice that to achieve an SNR of $\sim$10, we require an exposure time of 4 ks ($\sim$1 hour) and  a telescope time of $\sim$6 hours. The telescope time will be even longer for the distant sources than TW Hya.  Therefore, we are of the view that intra-day spectroscopic variability will be difficult to detect with the UVIT even for a TTS with a higher accretion rate.

In the bottom panel of \autoref{UVSpec}, we compare the UVIT spectrum with multi-epoch HST spectra after degrading to UVIT resolution. 
The comparison shows that the variation in continuum and line flux are prominent even in UVIT resolution. \autoref{table_line_flux} compares the line fluxes in different epochs. We noticed that C~{\sc iv} line flux varies from $\sim$0.5 -- $\sim$2 times of that measured with UVIT observation.  
Hence, we can study the inter-day accretion variability in TTSs using UVIT FUV spectroscopy. However, as discussed above a longer exposure time and even longer telescope time will be required for distance sources or the sources with higher extinctions.    
Therefore, UVIT spectroscopic observations are restricted to brighter targets with strongly accreting or highly active non accreting sources.


\section{Summary}
\label{summary}

We present the first UVIT/FUV spectrum of a TTS, TW Hya. TW Hya is also observed in NUV bands with UVIT, but in the photometric mode. The UVIT/FUV spectrum matches well with the low-resolution IUE spectrum. Combining the IUE observations with the UVIT, we found that the line flux, as well as the continuum, are variable in FUV and can be detected with low spectral resolutions. This indicates that the variable nature of the mass accretion rate in TW Hya can be studied with UVIT's low resolution FUV Grating. 
Using the C~{\sc iv} line luminosities from IUE, HST and UVIT spectroscopic observations, we measured the range in accretion luminosity (0.04--0.24 L$_\odot$) and mass accretion rates (0.8--4.8 $\times$ 10$^{-8}$ $M_\odot/yr$) onto TW Hya.

We performed SED analysis of TW Hya using photometric observations to estimate stellar parameters (luminosity, temperature, radius, $\log\ g$), UV/accretion luminosity and its corresponding temperature. We also performed another SED analysis using GALEX observation, not including UVIT. We found that the accretion luminosity value using UVIT observation (0.031 $L_\odot$) is more than two times compared to that found using GALEX observation (0.014 $L_\odot$). This also supports the variable accretion nature of TW Hya. However, other parameters found to be similar using both the observations: $T_{eff}$ = 3900$\pm$50 K; $\log\, g = 4.0$, radius = 1.2$\pm$0.03 $R_\odot$, black-body temperature corresponding to accretion luminosity = 14100$\pm$25 K. From the SED analysis using photometric observations, we found the mass accretion rate ranges between 0.28--0.62 $\times 10^{-8}$~$M_\odot/yr$.

Though there are many studies on young stars/star-forming regions using UVIT photometry, the spectroscopy capabilities of UVIT have not been explored to their full potential, especially in the case of young stars. Comparison with multi-epoch IUE and HST spectra  suggests that UVIT can be used to study the inter-day spectroscopic variability of young stars, but restricted to bright and strongly accreting stars. Study of intra-day accretion variability is not feasible with UVIT. 
UVIT spectroscopy facility 
in conjunction with the HST observations, can provide an insightful view of the high-energy processes associated with star formation and provide complementary observations to the ULLYSES program. 
However, we recommend to carry out a more dedicated multi-epoch spectroscopic and multi-band photometric observation, to put better constraints on the UVIT's potential to study accretion variability. 
This work also highlights the need to develop future UV spectroscopic missions such as the Indian Spectroscopic Imaging Space Telescope \citep[INSIST;][]{Subramaniam_INSIST_2022} to study the accretion process associated with star formation.

\appendix

\section{UVIT spectrum of different orbit}\label{appendix_orbit_wise_spectrum}

\begin{figure}
    \includegraphics[width=\linewidth]{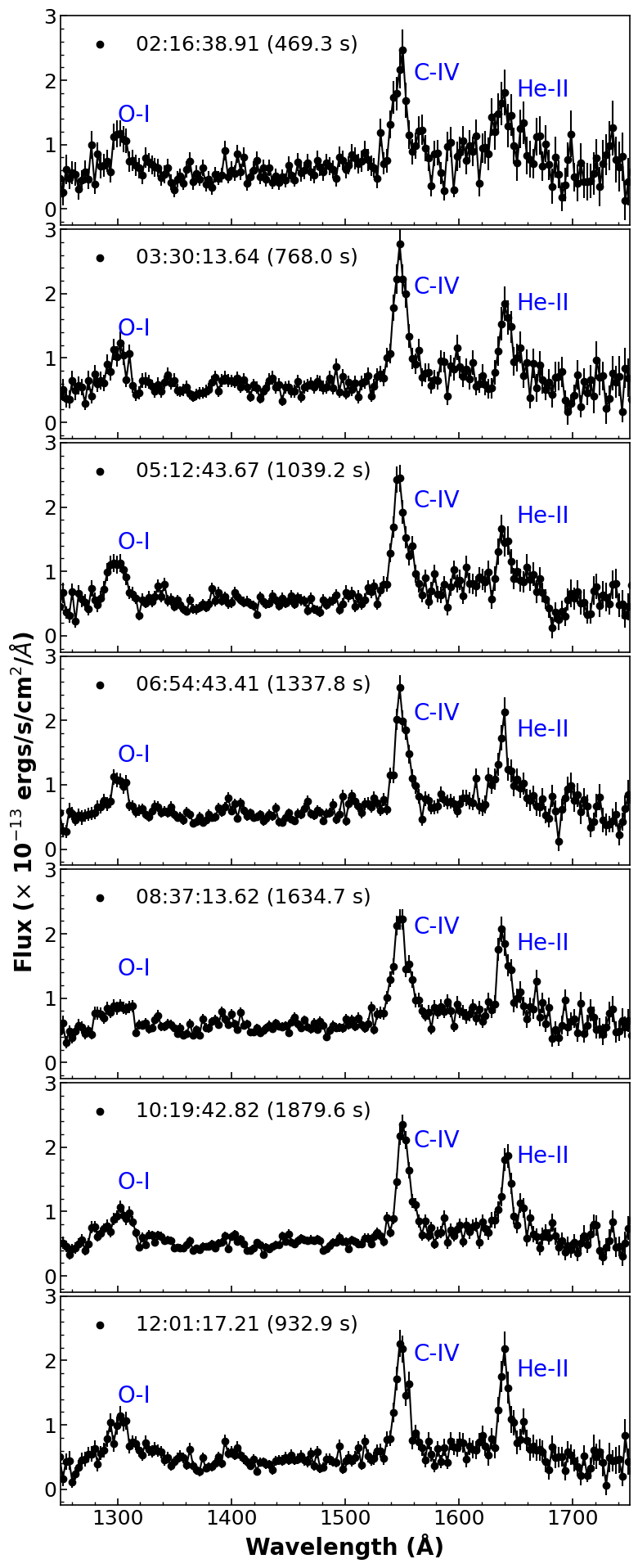}
    \caption{UVIT/FUV spectrum of TW Hya for different orbits. Start time of observation (exposure time) for each orbit is mentioned in the legend.}
    \label{diff_orbit_spectrum}
\end{figure}

\section*{Acknowledgements}

This publication uses UVIT data from the AstroSat mission of the ISRO, archived at the Indian Space Science Data Centre (ISSDC). The UVIT project is a result of collaboration between IIA, Bengaluru, IUCAA, Pune, TIFR, Mumbai, several centres of ISRO, and CSA. This publication uses UVIT data processed by the payload operations centre by the IIA. 
Based on observations obtained with the NASA/ESA Hubble Space Telescope, retrieved from the Mikulski Archive for Space Telescopes (MAST) at the Space Telescope Science Institute (STScI). STScI is operated by the Association of Universities for Research in Astronomy, Inc. under NASA contract NAS 5-26555. 
PKN acknowledges TIFR's postdoctoral fellowship. 
PKN also acknowledges support from the Centro de Astrofisica y Tecnologias Afines (CATA) fellowship via grant Agencia Nacional de Investigacion y Desarrollo (ANID), BASAL FB210003. 
DKO acknowledges the support of the Department of Atomic Energy, Government of India, under Project Identification No. RTI 4002. 

\vspace{-1em}



\bibliography{TW_Hya_UVIT}

\end{document}